\documentclass[a4paper,fleqn,usenatbib]{mnras}

\usepackage{newtxtext,newtxmath}
\usepackage[T1]{fontenc}
\usepackage{ae,aecompl}

\usepackage{graphicx}	% Including figure files
\usepackage{amsmath}	% Advanced maths commands
\usepackage{amssymb}	% Extra maths symbols
\usepackage[xindy]{glossaries}
\usepackage{tabularx}
\glsdisablehyper
\usepackage{subcaption}
\usepackage{dcolumn}
\newcolumntype{d}[1]{D{.}{.}{#1}}

\newcommand{\sups}[1]{\textsuperscript{#1}}
\newcommand{\subs}[1]{\textsubscript{#1}}

\makeglossaries

\newacronym{alfa}{ALFA}{Arecibo L-Band Feed Array}
\newacronym{dm}{DM}{Dispersion Measure}
\newacronym{frb}{FRB}{Fast Radio Burst}
\newacronym{fwhm}{FWHM}{Full-Width at Half-Maximum}
\newacronym{gbt}{GBT}{Greenbank Telescope}
\newacronym{htru}{HTRU}{High Time Resolution Universe}
\newacronym{igm}{IGM}{Intergalactic Medium}
\newacronym{ism}{ISM}{Interstellar Medium}
\newacronym{nip}{NIP}{Non-image Processing}
\newacronym{paf}{PAF}{Phased-Array Feed}
\newacronym{rfi}{RFI}{Radio-frequency Interference}
\newacronym{ska}{SKA}{Square Kilometre Array}
\newacronym{sefd}{SEFD}{System Equivalent Flux Density}
\newacronym{snr}{S/N}{Signal-to-Noise Ratio}
\newacronym{sps}{SPS}{Single Pulse Search}
\newacronym{vlbi}{VLBI}{Very Long Baseline Interferometry}

\title[The ALFABURST Commensal FRB Survey]{ALFABURST: A commensal search for
Fast Radio Bursts with Arecibo}

\author[G. Foster et al.]{Griffin Foster$^{1,2}$\thanks{E-mail: griffin.foster@physics.ox.ac.uk},
Aris Karastergiou$^{1,3,4}$,
Golnoosh Golpayegani$^{5,6}$,
\and Mayuresh Surnis$^{5,6}$, 
Duncan R. Lorimer$^{5,6}$,
Jayanth Chennamangalam$^{1}$,
\and Maura McLaughlin$^{5,6}$,
Wes Armour$^{7}$,
Jeff Cobb$^{2}$,
David H. E. MacMahon$^{2}$,
\and Xin Pei$^{8}$,
Kaustubh Rajwade$^{9}$, 
Andrew P. V. Siemion$^{2,10,11}$,
Dan Werthimer$^{2}$
\and and Chris J. Williams$^{1}$
\\
$^{1}$University of Oxford, Sub-Department of Astrophysics, Denys Wilkinson Building, Keble Road, Oxford, OX1 3RH,\\United Kingdom\\
$^{2}$Department of Astronomy, University of California, Berkeley, 501 Campbell Hall \#3411, Berkeley, CA, 94720, USA\\
$^{3}$Physics Department, University of the Western Cape, Cape Town 7535, South Africa\\
$^{4}$Department of Physics and Electronics, Rhodes University, PO Box 94, Grahamstown 6140, South Africa\\
$^{5}$Department of Physics and Astronomy, West Virginia University, Morgantown, WV 26505, USA\\
$^{6}$Center for Gravitational Waves and Cosmology, West Virginia University, Chestnut Ridge Research Building, Morgantown,\\ WV 26505, USA\\
$^{7}$OeRC, Department of Engineering Science, University of Oxford, Keble Road, Oxford, OX1 3QG, United Kingdom\\
$^{8}$Xinjiang Astronomical Observatory, Chinese Academy of Sciences, Urumqi, Xinjiang 830011, China\\
$^{9}$Jodrell Bank Centre for Astrophysics, University of Manchester, Oxford Road, Manchester M13 9PL, United Kingdom\\
$^{10}$Radboud University, Nijmegen, Netherlands\\
$^{11}$SETI Institute, Mountain View, California, USA\\
}

\date{Accepted XXX. Received YYY; in original form ZZZ}

\pubyear{2017}
\hypersetup{draft}
\begin{document}
\label{firstpage}
\pagerange{\pageref{firstpage}--\pageref{lastpage}}
\maketitle

% Abstract of the paper
\begin{abstract}
ALFABURST has been searching for Fast Radio Bursts (FRBs) commensally with other
projects using the Arecibo L-band Feed Array (ALFA) receiver at the Arecibo
Observatory since July 2015. We describe the observing system and report on the
non-detection of any FRBs from that time until August 2017 for a total observing
time of 518 hours.  With current FRB rate models, along with measurements of
telescope sensitivity and beam size, we estimate that this survey probed
redshifts out to about 3.4 with an effective survey volume of around
600,000~Mpc$^3$. Based on this, we would expect, at the 99\% confidence level,
to see at most two FRBs.  We discuss the implications of this non-detection in
the context of results from other telescopes and the limitation of our search
pipeline.  During the survey, single pulses from 17 known pulsars were detected.
We also report the discovery of a Galactic radio transient with a pulse width of
3~ms and dispersion measure of 281~pc~cm$^{-3}$, which was detected while the
telescope was slewing between fields.
\end{abstract}

% Select between one and six entries from the list of approved keywords.
% Don't make up new ones.

%%%%%%%%%%%%%%%%%%%%%%%%%%%%%%%%%%%%%%%%%%%%%%%%%%

\section{Introduction}
\label{sec:intro}

\glspl{frb} are short-duration, broad-band, dispersed pulses that are detected
at radio frequencies. They are mostly classified by virtue of their dispersion
being far in excess of the expected Galactic contribution. As for radio pulsars,
for FRBs, where we observe the pulse over a frequency band ranging from $\nu_1$
to $\nu_2$, the resulting dispersion delay 
\begin{equation}
\Delta t \propto {\rm DM} \, (\nu_1^{-2} - \nu_2^{-2}),
\end{equation}
where the \gls{dm} is the line integral of the electron column
density along the line of sight to the source.

Although the physical process that gives rise to FRBs is unknown, the
possibility that they originate at cosmological distances, and their potential
use as natural probes of large-scale structure and magneto-ionic content of the
Universe makes them worthy of attention. They appear as bright sources at the
telescopes on Earth, which indicates high luminosities given the implied
distance.  As short duration bursts, probably emanating from point-like sources,
they offer the unique opportunities to probe the inter-galactic medium
\citep[IGM;][]{2013ApJ...776..125M}, as pulsars do for the Galactic interstellar
medium.

Since the first reported detection \citep{2007Sci...318..777L}, a number of
surveys using a range of radio telescopes have attempted to detect further
bursts. At the time of writing, 25 \glspl{frb} have been reported \citep[for an
up-to-date list, see][]{2016PASA...33...45P}. While the majority of these have
been detected with the Parkes Radio Telescope at 1.4~GHz (L-band), other
telescopes are making important contributions. FRB~121102 was detected in the
Pulsar Arecibo L-band Feed Array (PALFA) \citep{2014ApJ...790..101S}. This
\gls{frb} is the only known \gls{frb} to repeat \citep{2016ApJ...833..177S}.
FRB~110523 was detected with the Green Bank Telescope (GBT) at 820~MHz
frequencies, confirming \glspl{frb} are observable outside L-band
\citep{2015Natur.528..523M}.  Recently, a number of very bright \glspl{frb} has
been detected with UTMOST at 843~MHz
\citep{2017MNRAS.468.3746C,atel10697,atel10867} and ASKAP at 1.4~GHz
\citep{2017ApJ...841L..12B}.

Even with the current small sample of FRBs population, it is clear that their
 properties vary significantly. The measured \glspl{dm} range from
176~pc~cm$^{-3}$ (FRB~170827) to 2596 pc~cm$^{-3}$ (FRB~160102), with pulse
widths ranging from sub-ms (unresolved) to 26~ms, and apparent flux densities
covering four orders of magnitude.  If the population is extragalactic then the
sky distribution is isotropic.  But, there is an apparent observational
disparity in the \gls{frb} event rate between high and low Galactic latitudes,
possibly due to diffractive interstellar scintillation
\citep{2015MNRAS.451.3278M}.

Single dish telescopes have been essential to detection of \glspl{frb} and
continue to be useful for population statistics.  But, these telescopes provide
limited localization.  The unknown detection position in the primary beam, and
one-off nature of most of the \glspl{frb} does also not allow precise
determination of the absolute flux density or the spectral index. Only the
repeater FRB~121102 has been localized using \gls{vlbi}
\citep{2017ApJ...834L...8M, 2017ApJ...834L...7T}. Localization is key to
understanding \glspl{frb}. This requires the use of interferometric arrays with
arc-second accuracy, such as MeerKAT, ASKAP, and the SKA. 

Apart from localization, \gls{frb} spectra offer important clues on the nature
of the emission process. Low frequency searches with LOFAR
\citep{2015MNRAS.452.1254K}, MWA \citep{2015AJ....150..199T}, and the GBT
\citep{2017arXiv170107457C} have reported non-detections.
A limited number of \gls{frb} surveys have been above L-band frequencies.
This is, in part, due to the narrowing of beam size which limits sky coverage.
V-FASTR \citep{2016ApJ...826..223B}, a commensal survey on the VLBA, has
reported a non-detection on observations up to 100~GHz.
\cite{2017arXiv170507553L} ran a coordinated-in-time, multi-telescope campaign
of the repeater \gls{frb}.  They report non-detection of pulses at VHF, C-band,
Ku-band during periods of detected bursts in L-band and S-band. \cite{atel10675}
report detections of FRB121102 from 4-8 GHz (C-band). In summary, our current
understanding of \glspl{frb} spectra is limited, however they appear not to
follow the steep power law example of radio pulsars, and may even not be smooth
and continuous with frequency.

For single dish telescopes there is a trade-off of sensitivity for survey speed.
Small dishes, such as those in the ATA `Fly's Eye' survey
\citep{2012ApJ...744..109S}, allow for a large sky coverage, but have low
sensitivity.  ASKAP dishes with \glspl{paf} provide a large sky coverage with a
significant enough sensitivity to detect bright \glspl{frb}. Conversely,
Arecibo provides the highest sensitivity, but with a very narrow beam.
The majority of \glspl{frb} have been discovered with Parkes using the
multi-beam system. The high sensitivity, large number of survey hours, and
increased field of view from using multiple beams all contribute to the large
number of detections. Interferometric arrays such as CHIME and MeerKAT will
provide both sensitivity and sky coverage. One important question relating to
the nature of the \gls{frb} population is what are the statistics of source
numbers versus source flux density, and whether or not the cumulative flux
density distribution is consistent with a population of cosmologically
distributed standard candles. To answer this question, it is particularly
interesting to sample both extreme ends of the flux density axis: the brightest
\glspl{frb} discovered using small telescopes in long duration and large
sky-coverage surveys, as well as the weakest \glspl{frb} sampled through
high-sensitivity observations with large telescopes, necessarily sacrificing
survey time and sky coverage. 

In this paper, we describe results from the ALFABURST survey, which has enabled
high sensitivity observations to better sample the low flux density end of the
population. ALFABURST makes use of the large amount of time spent by the ALFA
receiver for other astronomical surveys.  In Section \ref{sec:overview}, we
summarize the survey parameters and observations carried out so far.  A
wide-feature, learned model was used to classify each dataset in order to
filter out radio-frequency interference and create a priority queue for visual
examination. This model and the post-processing procedures are discussed in
Section \ref{sec:event_classify}.  Although no FRBs have been found in
observations carried out so far, we did detect one pulse that is consistent
with an origin in the Galactic plane. This source is discussed in Section
\ref{sec:18062017}. We discuss the expected event rates in Section
\ref{sec:event_rates}. Finally, in Section \ref{sec:discuss}, we consider
possible explanations for our non-detection of FRBs so far and speculate on
future developments.

\section{Observations}
\label{sec:overview}

\subsection{ALFABURST description}

ALFABURST is an \gls{frb} search instrument which has been used to commensally
observe since July 2015 with other \gls{alfa} observations at the Arecibo
Observatory. This system is a component of the SETIBURST back-end
\citep{2017ApJS..228...21C} and uses ARTEMIS \citep{2015MNRAS.452.1254K} for
automated, real-time pulse detection. We perform inline radio-frequency
interference (RFI) removal, baselining using zero-DM removal
\citep{2009MNRAS.395..410E}, and spectrum normalization before single pulse
detection. During this time period a \gls{sps} was performed from \gls{dm} 0 to
10000~pc~cm$^{-3}$, pulse widths from $256~\mu s$ to $16$~ms (using a logarithmic
decimation factor $D=1,2,4,\ldots,64$), across a 56~MHz bandwidth for
all 7 beams. We return to the effective \gls{dm} of the search in Section
\ref{sec:event_rates}. The gain of Arecibo allows for the most sensitive
\gls{frb} search to date.

Detections above a peak signal-to-noise ratio (S/N) of 10 were recorded along
with an $8.4$~s dynamic spectrum window around the event. When multiple events
were detected in the same time window, these events were pooled together and
recorded to disk.  Approximately $2.5 \times 10^5$ 8.4~s datasets were recorded
between July 2015 and August 2017, the vast majority of which are false
detections due to \gls{rfi} signals passing the real-time \gls{rfi} exciser. We
have detected no \glspl{frb} in our commensal survey.

%%% INLINE RFI EXCISION BEGINS %%%

\subsection{Inline RFI Excision}
\label{sec:rfi_excise}

An inline RFI exciser is implemented in the pipeline to mitigate strong RFI
sources. This leads to a significant reduction in the number of false-positive
detections in the dedispersion search.  Individual frequency channels in a
spectrum are replaced when the power exceeds a threshold $T_{\textrm{chan}}$
after the spectrum is normalized to zero mean, unity standard deviation
($\mu=0$, $\sigma=1$). Entire spectra are also clipped when their
frequency-integrated power exceeds a threshold $T_{\textrm{spectra}}$. For
standard ALFABURST operation $T_{\textrm{chan}} = 5$ and $T_{\textrm{spectra}} =
10$.  The \gls{rfi} exciser operates on data prior to any time decimation and 
integration ($D=1$).

For very bright pulses, the \gls{rfi} exciser will erroneously replace channels
or spectra, reducing the overall flux.  For the sensitivity of the \gls{alfa}
receiver, individual channels with flux greater than 2.8~Jy and,
frequency-integrated flux greater than $\sim250$ mJy are excised. The peaks of
bright \glspl{frb} such as FRB~150807 and FRB~170827 would be significantly
clipped by the exciser. But, the edges of the pulse would not. Both of these
\glspl{frb} would still be detected at a significant peak S/N. All previously
reported \glspl{frb} would be detected with ALFABURST at high peak S/N even if
partially clipped.

The zero-DM removal and spectral replacement affects low-DM pulses. For
reference, the minimum \gls{dm} before the total dispersive delay across the
band is equal to a time sample is \gls{dm}~$=1.8$~pc~cm$^{-3}$ for the
typical ALFABURST observing band (using Eq. 5.1 of \cite{2004hpa..book.....L}).
The minimum \gls{dm} before the dispersive delay within a single channel equals
the sampling time (also known as the diagonal \gls{dm}) is 
\gls{dm}~$=976$~pc~cm$^{-3}$. Single pulses from low-DM pulsars such
as B0834+06 are often clipped by the exciser, but are still detected at
significant peak S/N (see Table \ref{tab:knpsrtab}). As ALFABURST is focused on
detecting high-DM pulses, spectral replacement does not affect the survey
sensitivity.

%%% INLINE RFI EXCISION END %%%

%%% SYSTEM VERIFICATION BEGINS %%%

\subsection{Single Pulse Search Verification}
\label{sec:system_verify}

\begin{figure}
    \includegraphics[width=1.0\linewidth]{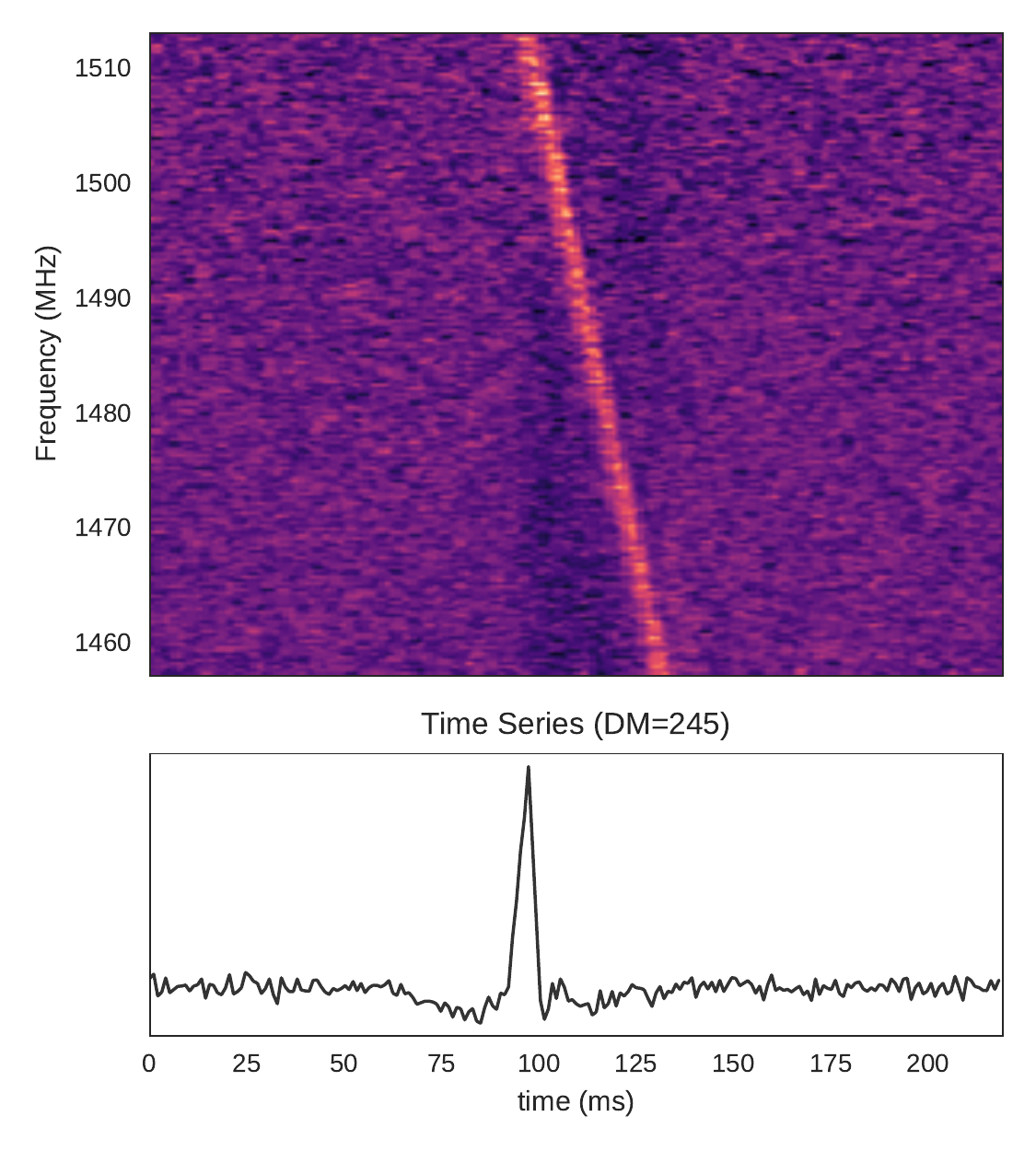}
    \caption{Detection of a single pulse from PSR B1900+01 (DM 245~pc~cm$^{-3}$). The
    baseline dip before and after the pulse is due to zero-DM removal
    \citep{2009MNRAS.395..410E}. }
    \label{fig:B1900}
\end{figure}

The PALFA survey schedule includes regular observations of known pulsars to
verify their data analysis pipeline. This provides a consistent verification of
our \gls{sps} to detect dispersed pulses. As the PALFA survey is targeted at the
Galactic plane, a number of high \gls{dm} pulsars were observed. Single pulses
from B1859+03 (\gls{dm}: 402), B1900+01 (\gls{dm}: 245) (Figure
\ref{fig:B1900}), B2002+31 (\gls{dm}: 234), B1933+16 (\gls{dm}: 158), among
others were detected. Table \ref{tab:knpsrtab} lists the parameters for the
known pulsars detected by the \gls{sps}.

\begin{table}
    \begin{center}
    
    \begin{tabular}{ld{2.1}d{3.2}d{3.0}d{4.0}d{2.1}}
    
    \hline
    \multicolumn{1}{c}{PSR} & \multicolumn{1}{c}{S\subs{1400}} & \multicolumn{1}{c}{DM\subs{cat}}    & \multicolumn{1}{c}{DM\subs{obs}}    & \multicolumn{1}{c}{N\subs{pulses}}   & \multicolumn{1}{c}{S/N\subs{max}} \\
                             & \multicolumn{1}{c}{(mJy)}        & \multicolumn{1}{c}{pc cm\sups{--3}} & \multicolumn{1}{c}{pc cm\sups{--3}} &                       &     \\
    \hline
    
    B0525+21    &   9.0  &   50.87  &   50  &     1  &  72.3  \\
    B0540+23    &   9.0  &   77.70  &   77  &     1  &  11.7  \\
    B0611+22    &   2.2  &   96.91  &  101  &  5192  &  48.8  \\ 
    J0631+1036  &   0.9  &  125.36  &  125  &     7  &  10.2  \\
    B0834+06    &   4.0  &   12.86  &    9  &   223  &  35.0  \\
    B1133+16    &  32.0  &    4.84  &    7  &   291  &  15.5  \\
    B1737+13    &   3.9  &   48.67  &   46  &  1880  &  49.4  \\
    B1859+03    &   4.2  &  402.08  &  402  &     2  &  20.4  \\
    B1900+01    &   5.5  &  245.17  &  246  &   151  &  35.4  \\
    J1908+0457  &   0.9  &  360.00  &  352  &     3  &  12.9  \\
    J1908+0500  &   0.8  &  201.42  &  202  &   160  &  18.5  \\
    J1910+0728  &   0.9  &  283.70  &  288  &     2  &  10.2  \\
    J1913+0904  &   0.2  &   95.30  &   97  &  1524  &  44.7  \\
    B1913+10    &   1.3  &  241.69  &  245  &     2  &  16.1  \\
    B1933+16    &  42.0  &  158.52  &  154  &    10  &  30.5  \\
    B1937+24    &    *   &  142.88  &  146  &    37  &  24.6  \\
    B2002+31    &   1.8  &  234.82  &  250  &     4  &  27.6  \\ 
    
    \hline
    
    \end{tabular}
    \end{center}
    
    \caption{Parameters for known pulsars detected in the ALFABURST survey. The
    columns from left to right are, pulsar name, mean flux density at 1400~MHz,
    catalog DM, observed DM of the strongest pulse, number of detected
    single-pulses, and maximum single-pulse S/N. The mean flux density at 1400
    MHz and catalog DM were obtained from the ATNF pulsar catalog (version
    1.56).}
    \label{tab:knpsrtab}
\end{table}

% alfaburst-initial-survey/notebooks/B1900_01.ipynb

%%% SYSTEM VERIFICATION ENDS   %%%

%%% SKY COVERAGE BEGINS %%%

\subsection{Survey Coverage}
\label{sec:survey_coverage}

\begin{figure*}
    \includegraphics[width=1.0\linewidth]{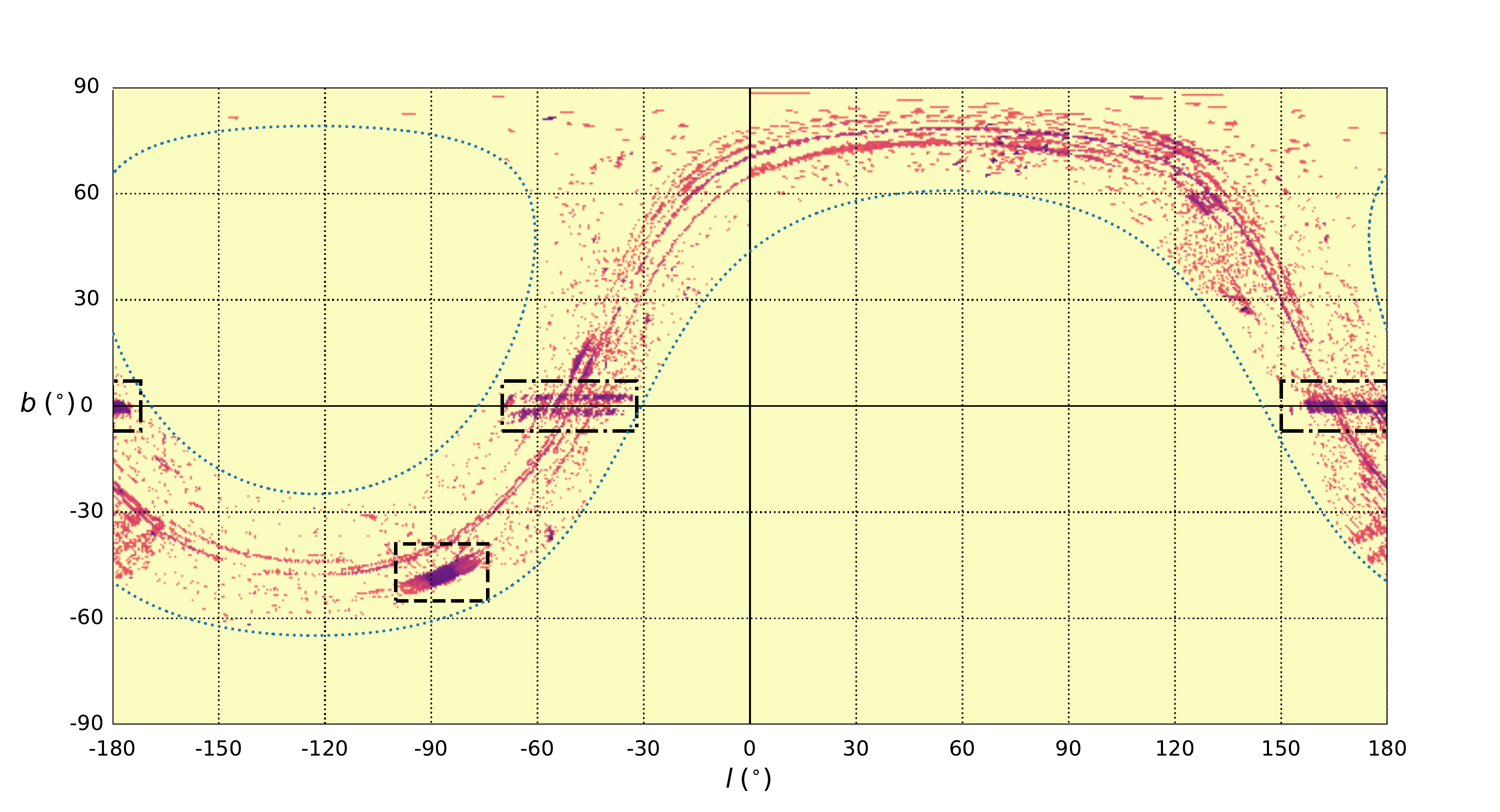}
    \caption{Sky coverage during ALFA usage between July 2015 and June 2017,
    shown in a Cartesian projection in Galactic coordinates along with
    declination pointing limits (blue dashed). Color represents total time
    pointing in a log scale. The majority of ALFA usage during this time was for
    the PALFA survey along the Galactic Plane (dot-dashed boxes) and the AGES
    survey (dashed box).  The S-shaped arcs across the plot are due to fixed
    pointings in local azimuth and altitude.
    }
    \label{fig:sky_coverage}
\end{figure*}

Since ALFABURST was installed, the majority of ALFA observation time is
allocated for the AGES \citep{2006MNRAS.371.1617A} and PALFA
\citep{2006ApJ...637..446C} surveys (Figure \ref{fig:sky_coverage}).  The AGES
survey pointings are off the Galactic Plane, which is ideal for \gls{frb}
surveys.. PALFA is a pulsar search survey with pointings near the Galactic
Plane. These lines of sight can introduce significant dispersion due to the
\gls{ism}. We search out to a DM of 10$^{4}$~cm$^{-3}$~pc which is well beyond
the maximum Galactic dispersion, but within the technical capabilities of our
system. 

% alfaburst-initial-survey/notebooks/Sky_Coverage.ipynb

Approximately 65\% of the ALFABURST survey time has been in pointings out of the
Galactic Plane ($|b| > 5^{\circ}$).  These pointings are primarily from the
ongoing AGES survey.  Pointings in the plane are primarily from the PALFA
survey.  The PALFA survey detected the repeating \gls{frb} FRB121102
\citep{2014ApJ...790..101S}, the only \gls{frb} detected with Arecibo thus far.
As ALFABURST has been running commensally with the PALFA survey since 2015 these
two back-ends act as independent single-pulse search pipelines, useful for
detection verification.  Since the beginning of ALFABURST observations no
\glspl{frb} have been reported by PALFA. No follow-up observations of FRB~121102
have been conducted using ALFA.

%%% SKY COVERAGE ENDS   %%%

%%% OBSERVATION TIME BEGINS %%%

\subsection{Observing Time}
\label{sec:obs_time}

From the beginning of July 2015 to the end of April 2017 \gls{alfa} has been
used for approximately 1400 hours of observing, with all seven beams functional.
Due to pipeline development and hardware reliability, ALFABURST was active and
functional for, on average, 322 hours per beam.  The current system is set up to
be reliably in use for all beams any time \gls{alfa} is active and in the
correct receiver turret position. Since April 2017 this stable version of the
pipeline has run for an additional 196 hours. This has resulted in a total of
518 hours of processed observing time since ALFABURST began commensal observations.

%%% OBSERVATION TIME ENDS   %%%

%%% PRIORITIZER BEGINS %%%

\section{Event Classification Strategy}
\label{sec:event_classify}

The significant DM trial range, variety of \gls{rfi} events, and commensal
nature of the survey, leads to a large number of false detections.
Approximately $2 \times 10^5$ unique 8.4~s datasets were recorded with at least
one detection above the minimum peak S/N threshold of 10. In order to reduce the
number of events we need to visually inspect we have developed a prioritizer
model based on a trained probabilistic classifier. The use of trained
classifier models is becoming a common post-processing technique in \gls{frb}
surveys \citep{2016PASP..128h4503W} in order to manage the large number of
detected events. Our model can be found in the survey git
repository\footnote{https://github.com/griffinfoster/alfaburst-survey}.

Building the model involved inspecting and labelling a sample of the events. We
used a sample set of approximately 15,000 event windows.  For each event window,
a diagnostic plot was generated which contained the original dynamic spectrum,
the dedispersed dynamic spectrum of the S/N-maximized DM, along with a frequency
collapsed time series of the detection.  During figure generation 409 features
were also computed to be used in the model.  These features include statistics
such as the number of triggers in the event window, the DM range of these
triggers, and the median, mean, and standard deviation of a coarse pixelization
of the dynamic spectrum ($4 \times 16$) and S/N maximized dedispersed time
series (16 segments). A complete list of the features can be found in the
survey git repository. These raw features were reduced during model
pre-processing to 398 features.

In order to build a classifier model using the derived event statistics, a
sample of events were visually inspected and labelled into 8 classes of RFI,
systematic effects, and astrophysical source (pulsars) (Table
\ref{tbl:event_classes}). These heuristic classes were based on multiple,
iterative inspections of the sampled events. A simple binary astrophysical
classifier of events leads to a poor model because the types of events which
are non-astrophysical take on a variety of forms.

\begin{table}
\centering
\begin{tabularx}{\linewidth}{clX}
\hline
Class ID & $N_{\textrm{events}}$ & Description                                     \\
\hline
1        & 151     & Unclipped low-level RFI                                       \\
2        & 4159    & Wide-band, duration \textgreater 1 second clipped RFI (2016+) \\
3        & 1898    & Wide-band, duration \textless 1 second clipped RFI (2016+)    \\
4        & 448     & Wide-band, short duration clipped RFI (2015)                  \\
5        & 617     & Sharp bandpass transition                                     \\
6        & 4649    & Wide-band, bursty, clipped RFI (2015)                         \\
7        & 863     & Error in spectra capture or replacement                       \\
8        & 1594    & Systematic int/float overflow                                 \\
9        & 691     & Astrophysical pulse or unknown event                          \\
\hline
Total    & 15070   &                                                              
\end{tabularx}
\caption{Event classes and distribution from the sample of labelled events used
to train the priority classifier model.}
\label{tbl:event_classes}
\end{table}

The class distribution is time-dependent as the detection pipeline has been
updated, the RFI environment has changed, and the telescope observing schedule
has changed over the time the survey has run. Classes 2 and 3 occur after the
inline RFI exciser was improved in 2016. Whereas classes 4 and 6 are events that
occur with the original RFI exciser. Because the ALFABURST is operating in
commensal mode, the band can unexpectedly be changed due to a change in the
observing frequency, these event windows are labelled as class 5 events. Class 7
and 8 are due to packet loss and incorrect digital gain settings. We found that
class 8 events can be removed simply by checking for overflow values in the
spectra, and therefore this class is dropped before building a classifier model.

Pulses from known pulsars were used as a proxy class for the FRB class. The
number of astrophysical pulse detections was low compared to the total number of
false-positive detections. It was necessary to use a large number of classes
as RFI and systematic effects took on a variety of forms.  This had the
additional effect of balancing out the number of events in each class, making
model training more robust.

These features along with the labels were used to build a random forest
probabilistic classifier model \citep{Ho:1995:RDF:844379.844681,Breiman2001}
using the \texttt{scikit-learn} package \citep{scikit-learn}.
This model is then used to probabilistically predict which class belongs to an
unlabelled data set. A one vs. the rest multi-class classifier strategy was used
for training. Before training, the features of the labelled data sets were
median removed and standard deviation normalized using an interquartile robust
scaler.  A random forest of 80 trees and 20 random features per node split was
found to produce the best score in a hyper-parameter grid search using a
log-loss scoring metric. During training and hyper-parameter optimization a
stratified k-fold cross-validation (3 splits)
\citep{Kohavi:1995:SCB:1643031.1643047} procedure was used.

\begin{figure}
    \includegraphics[width=1.0\linewidth]{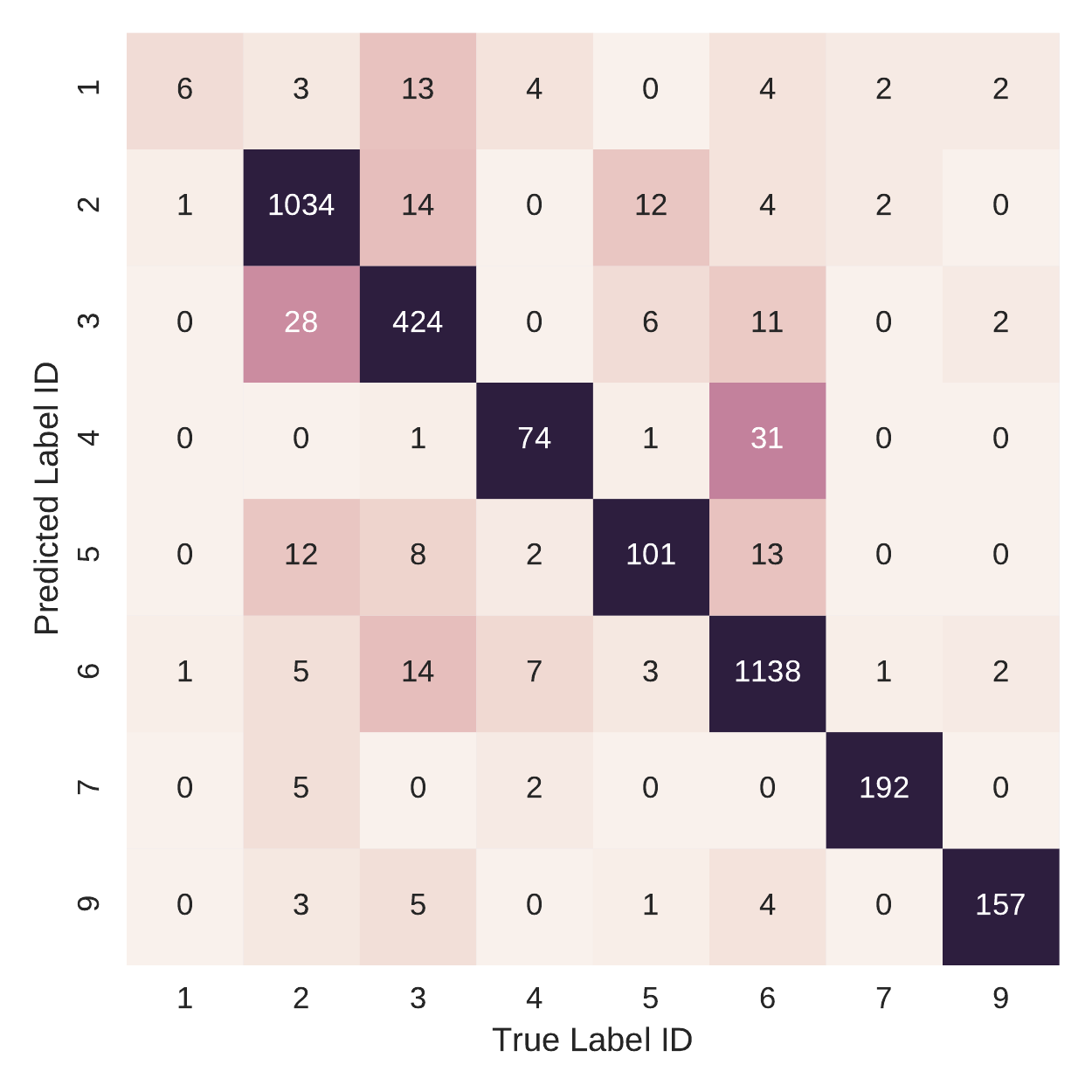}
    \caption{Confusion matrix of labelled testing data set after training the
    random forest model with the labelled training data set.
    }
    \label{fig:confuse}
\end{figure}

The trained model is successful at predicting the majority of the astrophysical
events to be astrophysical with high probability, as shown in the confusion
matrix (Figure \ref{fig:confuse}) when using 75\% of the labelled events for
training, and 25\% for testing. Of the non-astrophysical events, only 13 events
were misclassified as being likely astrophysical. A reasonably small number of
false-positive events to inspect. But, of the 163 astrophysical pulses in the
testing set, 6 events were misclassified. This is a more serious issue as we
would like to minimize the number of false-negative events for astrophysical
event windows.

% watermark:/home/griffin/data/alfa/priorityModel/notebooks/modelSandbox.ipynb

\begin{figure}
    \includegraphics[width=1.0\linewidth]{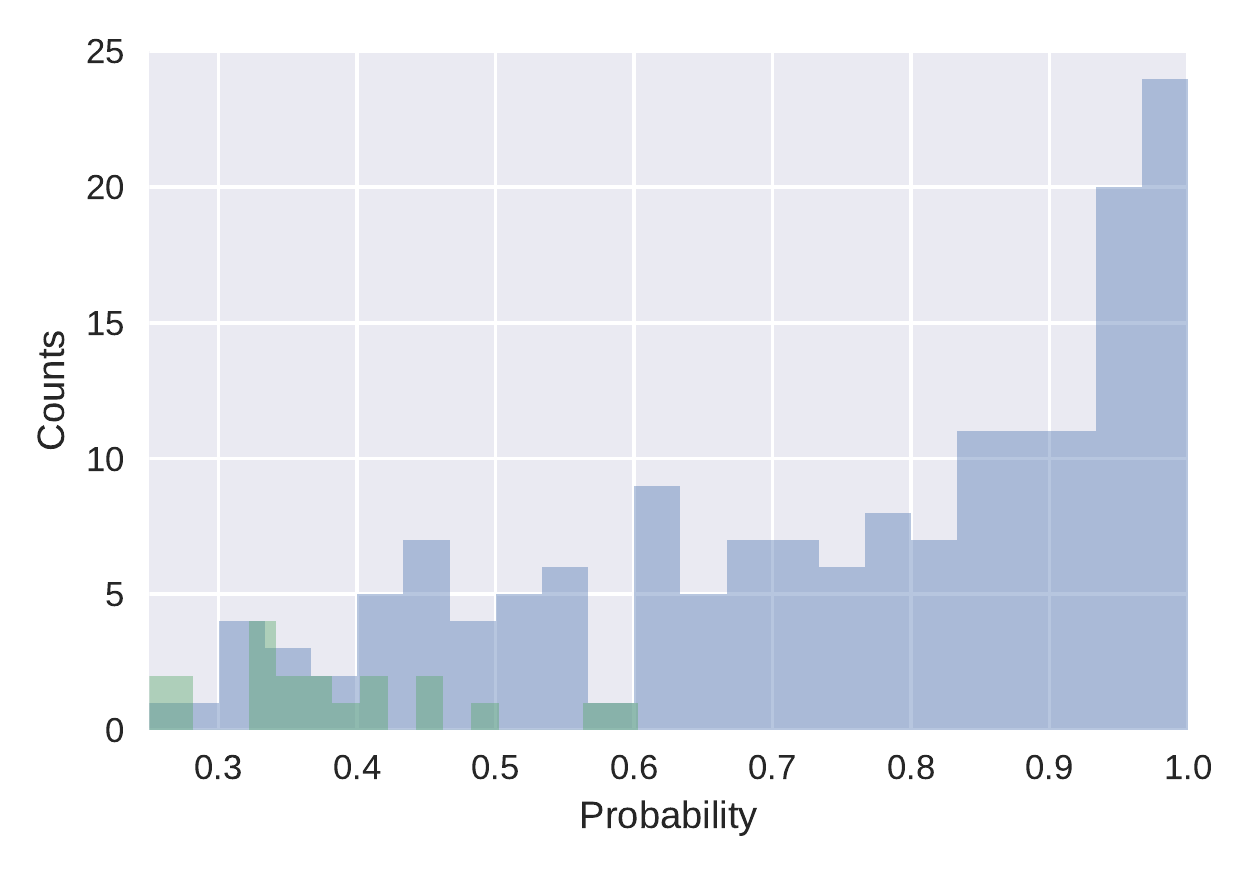}
    \caption{Histogram of the probability that a given astrophysical event
    (class 9, blue) and all other classes (green) is predicted to be an
    astrophysical event using our probabilistic classifier model.
    }
    \label{fig:class_hist}
\end{figure}

In searching for \glspl{frb} we are inclined to allow for a large number
of false-positive events (detection due to RFI or systematics) as long as there
are no false-negative events (pulses classified as RFI), i.e. a high recall for
astrophysical pulses. But, the confusion matrix is computed based on a discrete
class classification. The probabilistic predictions of all the astrophysical
pulses to be of the astrophysical class in the test set are all above 0.25
(Figure \ref{fig:class_hist}), while 20 events are reported as
false-positive for class 9 above this probability. We use this threshold to
select the top candidates from the survey. The events are sorted into a priority
queue based on the probability the event is astrophysical.

% watermark:/home/griffin/data/alfa/priorityModel/notebooks/modelSandbox.ipynb

Using a probabilistic multi-label classifier allows us to prioritize the order
and amount of time we spend on examining event datasets. Those with high
probability of belonging to a single class can be examined as a group quickly.
Datasets which fall into multiple classes are examined more thoroughly, they are
labelled by hand, and the set of features extracted during the figure generation
process is refined to further differentiate classes. This model building,
prioritizing, and examination process was iterated on multiple times to improve
the classifier. We continue to iterate on this model and will use it for future
prioritization of examining events.

We have not used the classifier model directly in our pipeline as the black-box
nature of the model can lead to misclassification, rather we have used it to
create a priority queue.  We have also used our classifier model as a data
exploration tool to add and refine procedural filters to the data.  An output of
the random forest model is the sort the features by 'importance' for
classification. For example, the most important feature for correctly
classifying a class 1 event (long duration replaced RFI) was the length of the
longest period of the dynamic spectrum with a derivative of zero. This makes
sense, as wideband RFI is replaced by a mean-zero noise spectrum. The most
important features for correctly predicting an astrophysical pulse were the
statistics from coarse pixelization of the dedispersed time series. This can be
attributed to the detection of a high S/N event in an otherwise noisy time
series.

Understanding the feature importance has led to the development of a
number of simple filters to reduce the number of false-positive detections
without relying on the classifier model. Data sets were cut if any of the
following criteria were met:

\begin{itemize}
    \item The maximum DM of events was less than 50~cm$^{-3}$~pc.
    \item Given the optimal dispersion measure, DM$_{\textrm{opt}}$, obtained
    from the S/N-maximized DM trial, if the DM range exceeds $(0.5 \times {\rm
    DM}_{\textrm{opt}}, 1.5 \times {\rm DM}_{\textrm{opt}})$, then the event is
    due to long duration RFI.
    \item More than 50\% of the spectra were replaced in the dataset.
    \item Any values in the dataset exceed the \texttt{int32} maximum value.
    These are is class 8 events, and due to errors in receiving packets.
\end{itemize}

These filters were applied to each dataset in post-processing to reduce the
number of datasets to approximately 30,000. The windows were sorted by S/N, and
the top S/N events were examined first.  During this process all datasets were
labelled.  Astrophysical events were identified based on the beam ID and
pointing information.

%%% PRIORITIZER ENDS %%%

%%% JUNE 18, 2017 BEGINS %%%

\section{The event of 2017, June 18}
\label{sec:18062017}

\begin{figure}
    \includegraphics[width=1.0\linewidth]{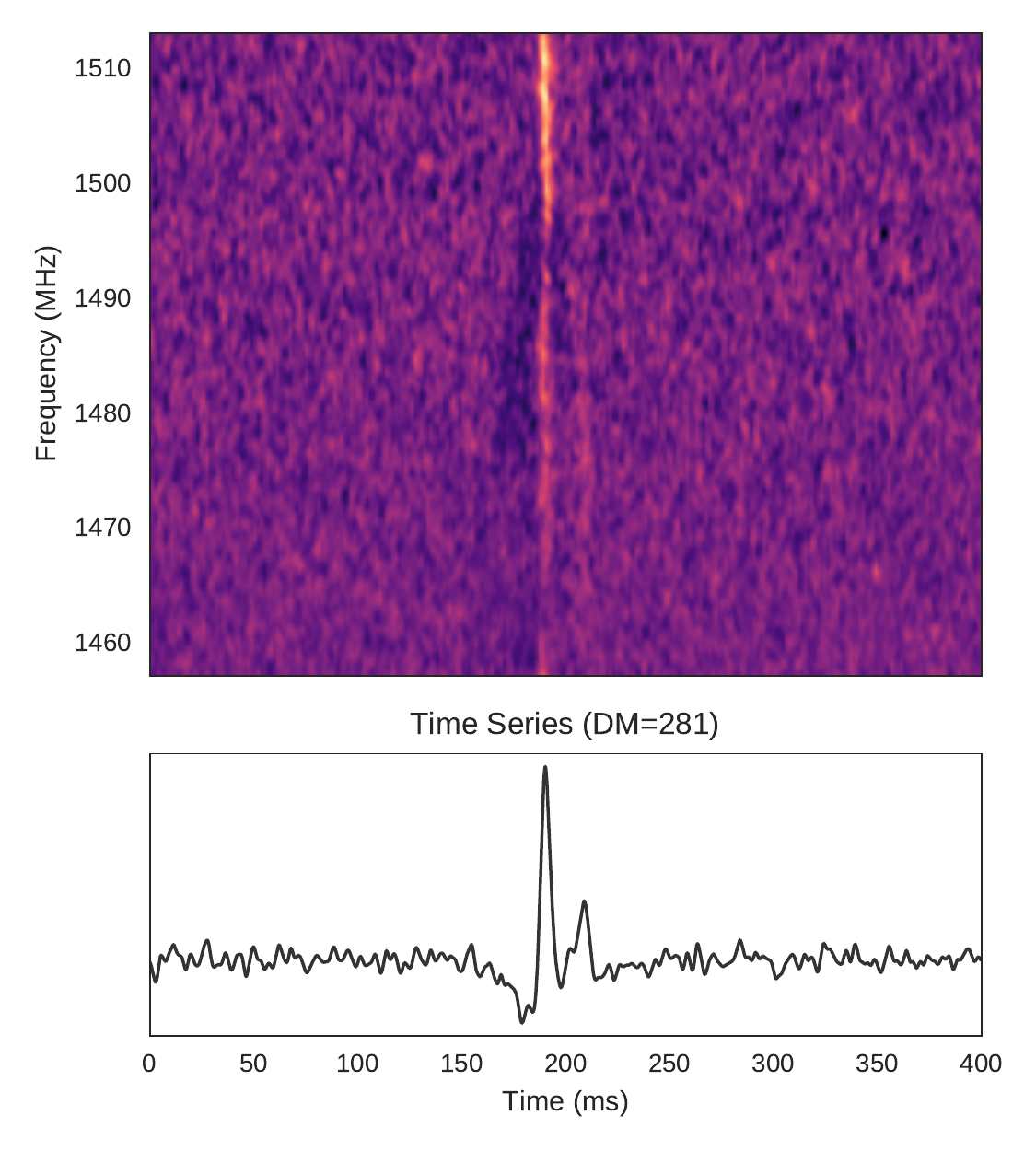}
    \caption{A broad band pulse (S/N maximized at DM~$=281$~pc~cm$^{-3}$)
    detected in beam 5 while the telescope was slewing during a PALFA
    observation. There is no known source which has been associated with this
    detection. As the observation was in the Galactic Plane it is likely
    Galactic in origin.
    }
    \label{fig:D20170618_spectrum}
\end{figure}

Though we report no detection of FRBs in the first two years of observations
with ALFABURST we have made an initial detection of an as yet unknown broad-band
(within our band) pulse (Figure \ref{fig:D20170618_spectrum}) at a peak S/N of
18. The peak S/N is maximized by dedispersion using a DM of 281~pc~cm$^{-3}$ and
time decimation factor 8. The main pulse width is approximately 3~ms. The pulse
occurred in beam 5, and there were no other detections in the other beams at the
time.

% watermark:/home/griffin/data/alfa/D20170618/Beam5_fb_D20170618T005616.buffer2-paper.ipynb

\begin{figure}
    \includegraphics[width=1.0\linewidth]{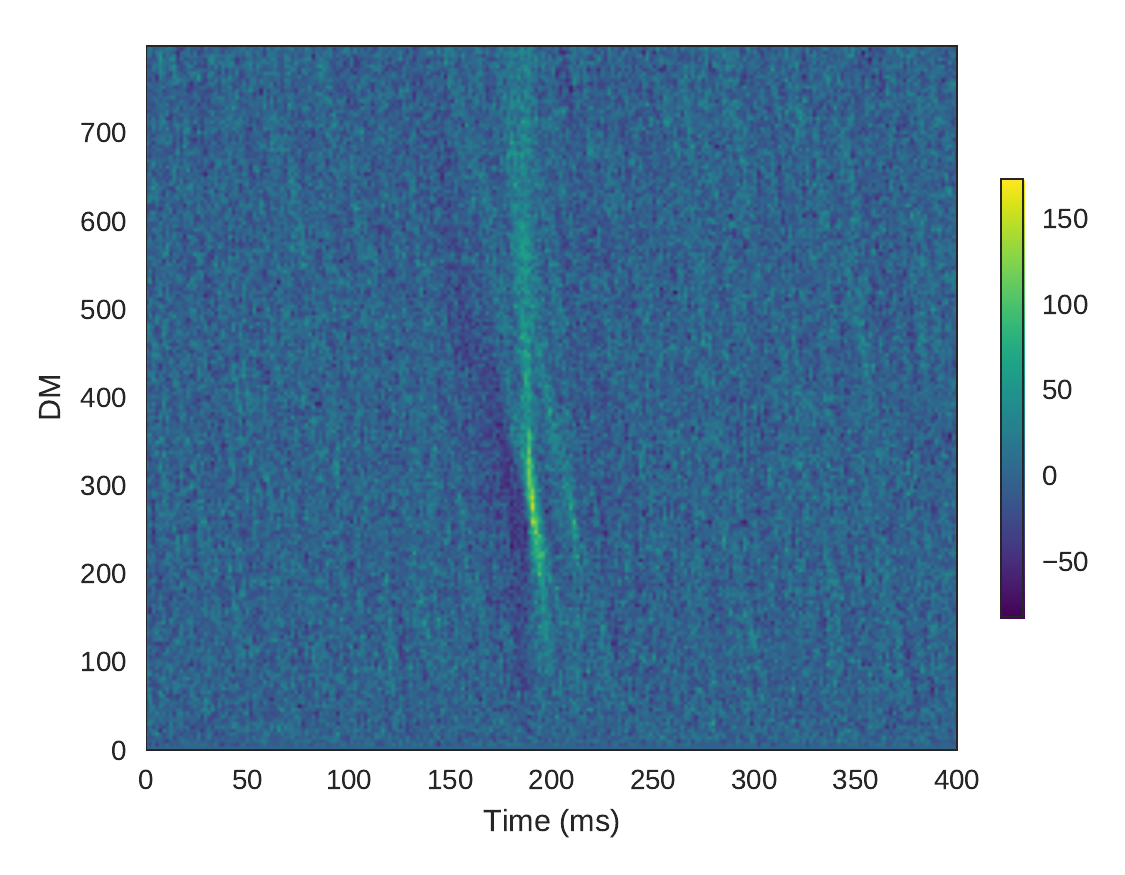}
    \caption{DM-time plot of the 2017 June 18 pulse. The pulse is compact in
    DM-time space, consistent with an astrophysical event. The secondary pulse
    20~ms after the primary pulse causes the intensity to be slightly elongated
    to higher trial DMs.
    }
    \label{fig:D20170618_dmspace}
\end{figure}

The pulse is made up of two clear components, with the secondary
pulse arriving approximately 20~ms after the primary pulse, as seen in the
dynamic spectrum (Figure \ref{fig:D20170618_spectrum}). In DM-time space the
event is compact, consistent with a $\nu^{-2}$ dispersion relation (Figure
\ref{fig:D20170618_dmspace}), though such a fit has large error bars due to the
small fractional bandwidth that is processed with ALFABURST.

% watermark:/home/griffin/data/alfa/D20170618/Beam5_fb_D20170618T005616.buffer2-paper.ipynb

The detection occurred at 04:56:16~UT on 2017, June 18 (MJD 57922) during a
PALFA observing run. The event was not seen by the PALFA collaboration as it
occurred when the telescope was slewing between fields and the PALFA
spectrometers were not running. This is the first known detection of a
transient, broad-band pulse using ALFA during such a slew. However, this makes
it challenging to determine the accurate source position. Pointing information
from Arecibo is reported every second.  During the detection the pointing was
changing by approximately $5'$ per second in right ascension $2'$
per second in declination. This rate gives us a conservative estimate
of the error in pointing at the time the pulse was detected. Based on the time
stamp of the pulse and the pointing data the pulse occurred when beam 5 of ALFA
was pointing at right ascension: 18~h 45~m $10 \pm 20$~s, and declination: +00 d
$38 \pm 2$' (Galactic coordinates $l: 32.78 \pm 0.05^{\circ}, ~b: +1.68 \pm
0.05^{\circ}$).

This beam 5 pointing is close to the Galactic plane in the first quadrant. The
DM distance estimated from the NE2001 model \citep{2002astro.ph..7156C} is
approximately 6 kpc, which is well within the Galaxy. The maximum Galactic
contribution along this line of sight would produce a DM of
$\sim800$~pm~cm$^{-3}$. A search of the ATNF pulsar
database\footnote{http://www.atnf.csiro.au/people/pulsar/psrcat}
\citep{2005AJ....129.1993M}, Rotating Radio Transient (RRAT)
catalog\footnote{http://astro.phys.wvu.edu/rratalog}, and recent PALFA
discoveries \footnote{http://www.naic.edu/$\sim$palfa/newpulsars/} revealed no
known source with a DM near 281 pc~cm$^{-3}$ within a degree of the pointing.

As the telescope was slewing at the time, the source was only in the primary
lobe for a fraction of a second (assuming it was in the primary lobe and not a
side lobe). A source on the edge of the \gls{fwhm} beam would transit the beam
in a maximum of 500~ms for the slew rate of the telescope (at the ALFABURST
observing frequency this corresponds to a dispersed pulse with a maximum DM of
3500~pc~cm$^{-3}$).  It could therefore be a RRAT which we serendipitously
detected at the correct moment, or it could be an individual pulse from a
pulsar. This event is similar to FRB010621 \citep{2011MNRAS.415.3065K}
which is likely of galactic origin \citep{2014MNRAS.440..353B}.
\cite{2012MNRAS.425L..71K} suggest FRB010621 is due to pulsar giant pulse or
annihilating black holes. The second component would seem to rule out the latter
interpretation in this instance. This region has been previously surveyed with
PALFA and the Parkes Multi-beam Survey \citep{2001MNRAS.328...17M} with no
significant detection of a pulsar at this DM.

The pulse appears brighter at higher frequencies, which could be due to
scintillation. Another reason for this frequency-dependent structure is that the
pointing of the telescope is changing during the total dispersion time of the
pulse within the observed band. As the pointing moves, the corresponding
telescope gain also changes.  There was a higher beam gain at the beginning of
the pulse compared to the end of the pulse, inducing a frequency-dependent gain
response due to the beam, also known as \emph{spectral colorization}.  A more
detailed analysis of this event and the results of follow-up observations, will
be presented elsewhere.

%%% JUNE 18, 2017 ENDS %%%

%%% EVENT RATES BEGINS %%%

\section{Expected FRB Events}
\label{sec:event_rates}

The currently known 25 FRBs vary significantly in \gls{dm}, pulse width, and
flux density. Despite this, we assume a simple model to derive an expected event
rate with our survey\footnote{Jupyter notebooks used to carry out this work are
freely available and are hosted at
https://github.com/griffinfoster/alfaburst-initial-survey}.  We use a model
\citep[see equation 9 of][]{2013MNRAS.436L...5L} which assumes \gls{frb} sources
are standard candles with a fixed spectral index, uniformly distributed in
co-moving volume. The event rates in this model have been updated to the
event rates reported in \cite{2016MNRAS.460.3370C}. For an observed pulse of
typical width 4~ms (see below), these event rates are in the range 1100--7000
bursts per sky per day, where the range indicates statistical errors for the
99\% credible region.

Taking advantage of the large forward gain of Arecibo, we account for the
sensitivity of the 7 \gls{alfa} beams out to the outer edge of the first side lobe.
In practice we do this by splitting the beam and first side lobe into shells of
progressively lower gain but larger sky coverage, and integrate to obtain the
totals.  An \gls{alfa} beam is approximately 3.8'~$\times$~3.3' at \gls{fwhm}
across the band.  The ALFA beam is known to be relatively fixed in size
across the band due to the optics \citep{GALFAbeam}.  Given the average
observing time per beam of 518 hours this results in a survey coverage of $\sim
10 \; \textrm{deg}^2 \; \textrm{hours}$ when accounting for all 7 beams. This is
a small survey coverage compared to most other \gls{frb} surveys, primarily due
to the narrow beam size of Arecibo. The combined Parkes multi-beam surveys have
a total of 8231 observation hours \citep{2016MNRAS.460.3370C}, and a FWHM survey
metric of $\sim 4500 \; \textrm{deg}^2$ hours.  ALFABURST does not compete with
other surveys on sky coverage, rather it competes on sensitivity. This results
in probing a greater redshift range than for Parkes. Using Equation 6 of
\cite{2015MNRAS.452.1254K}, a  single-pulse-search pipeline is sensitive to
pulses with a minimum flux density
\begin{equation}
S_{\rm min} = \textrm{SEFD} \frac{\textrm{S/N}_{\rm min}}{\sqrt{D \; \Delta \tau \;
\Delta \nu}}
\end{equation}
which is a function of the telescope \gls{sefd}, the minimum S/N detection level
$\textrm{S/N}_{\rm min}$ and the decimation factor $D$ compared to the native
instrumental time resolution $\tau$, this comes from the search pipeline which
averages together spectra to search for scattered pulses. ALFABURST has a native
resolution of $\Delta \tau = 256 \; \mu s$, effective bandwidth $\Delta \nu = 56
\textrm{MHz}$, and $\textrm{S/N}_{min} = 10$. The FWHM \gls{sefd} of the
\gls{alfa} receiver is approximately 3~Jy across the band for all beams.

\begin{figure}
    \includegraphics[width=1.0\linewidth]{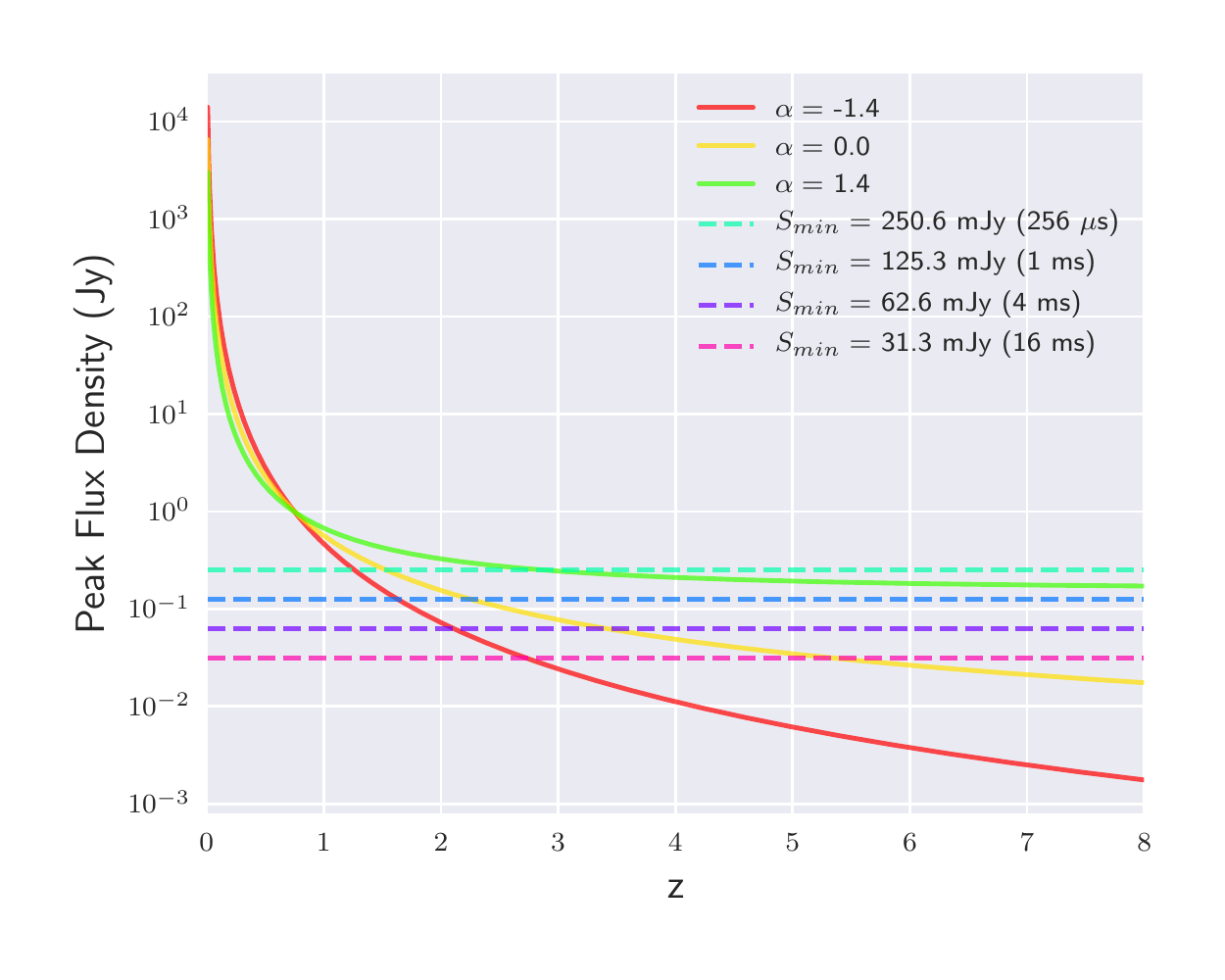}
    \caption{Sensitivity of the ALFABURST search pipeline (dashed) to FRB pulses
    assuming a standard candle model using different spectral index models
    (solid).
    }
    \label{fig:fwhm_sefd_z}
\end{figure}

The \gls{sps} pipeline is configured to search for pulses from 256~$\mu$s to 16
ms. Considering only the main beam lobe, a perfect matched filter would result
in a sensitivity to pulses with a minimum frequency-averaged flux of $S_{256
\mu\textrm{s}} = 250$ mJy to $S_{16 \; \textrm{ms}} = 31$~mJy
\citep{2015MNRAS.452.1254K}. Figure \ref{fig:fwhm_sefd_z} shows the peak flux
density of using the standard candle \gls{frb} model as a function of source
redshift for different model spectral indices. The dashed lines of constant flux
show the sensitivity of the ALFABURST search pipeline to pulses of different
widths. Assuming a positive spectral index model ($\alpha=1.4$) results in a
sensitivity out to the maximum redshift/\gls{dm} for pulses with widths of at
least 1 ms. A flat spectral index model results in sensitivity from $z \sim 1.5$
(256~$\mu$s) out to $z \sim 5$ (16~ms) depending on pulse width. A negative
spectral index model ($\alpha \sim -1.4$) limits the survey to $z < 3$ for all
pulse widths.

% alfaburst-initial-survey/notebooks/ALFABURST_Derived_FRB_Rates.ipynb

If we assume a simple model of $\alpha=0$ as we have limited information about
the source spectral index, and a pulse width of 4 ms as that is an approximate
median pulse width of reported \glspl{frb}, then this results in a maximum
redshift of $z=3.4$ (a co-moving distance of 6.8 Gpc) and a survey volume of $6
\times 10^5$ Mpc$^3$ when using all 7 \gls{alfa} beams. The number of galaxies
sampled in this volume is $6 \times 10^3$ assuming a constant galaxy number
density of $10^{-2}$ per Mpc$^3$.  The volumetric event rate from
\cite{2013Sci...341...53T}, is stated to be $R_{\textrm{FRB}} = 10^{-3}$
\glspl{frb} per galaxy per year. Adopting the more realistic lower rates found
by \cite{2016MNRAS.460.3370C} based on a larger sample of discoveries, we adopt
$R_{\textrm{FRB}}$ to be in the range $1.1 \times 10^{-4}$---$7.0 \times
10^{-4}$ \glspl{frb} per galaxy per year.  With these assumptions, we do not
expect any FRB detections based on the current observation time. We note once
again that the areal coverage used in this calculation is only based on the
sensitivity and size of the main beam lobe.

\begin{figure}
    \includegraphics[width=1.0\linewidth]{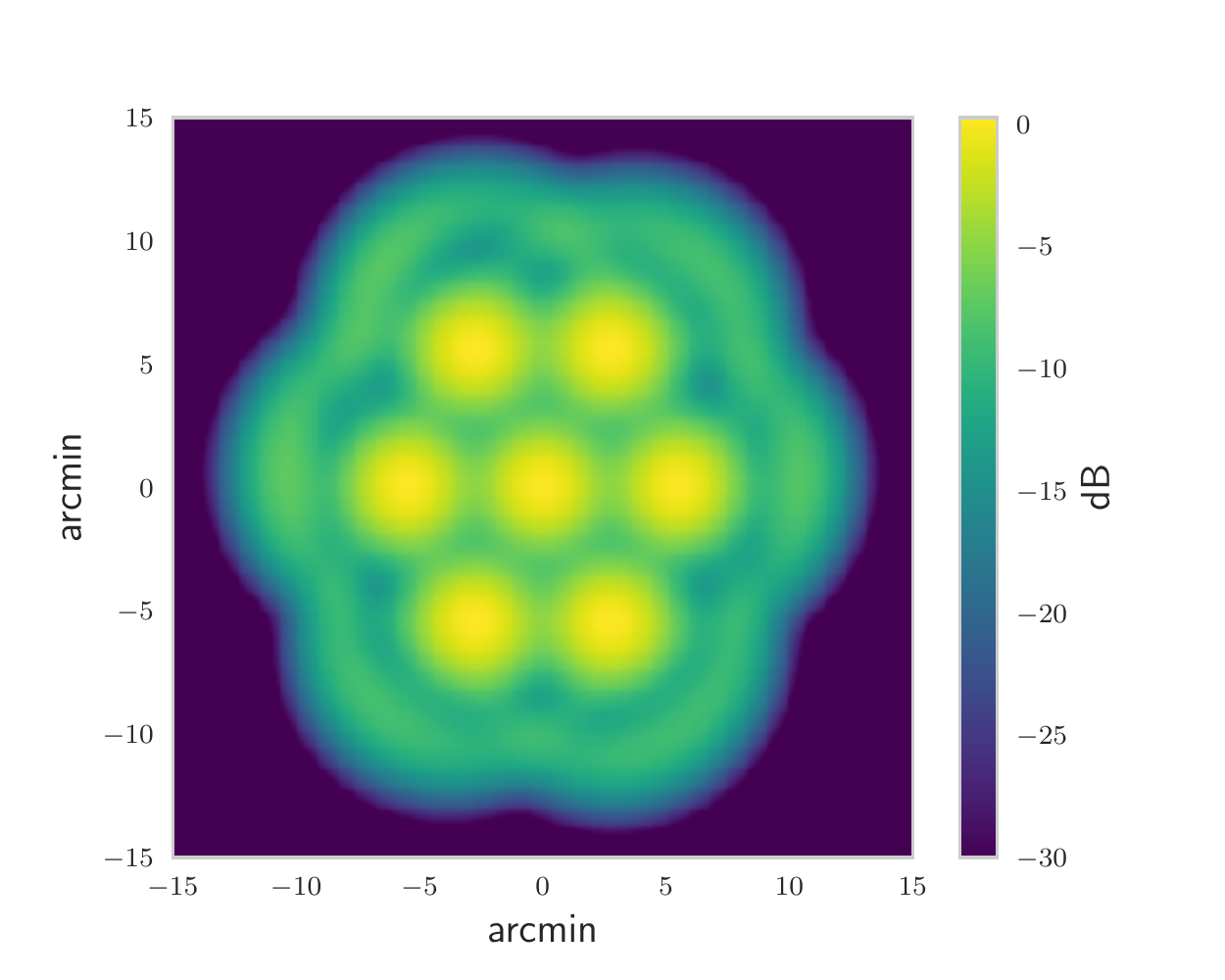}
    \caption{Primary and first side lobe model of the AFLA receiver in
    decibels, cut-off at $-30$ dB.The first side lobe peak at around $-9$ dB.
    }
    \label{fig:alfa_beam}
\end{figure}

As mentioned above, it is worth also taking into account the entire first side
lobes of the beams as Arecibo would be sensitive to detect most previous
\glspl{frb} in these. Using the parameterized \gls{alfa} beam model (Figure
\ref{fig:alfa_beam}) \citep{GALFAbeam} we can compute the \gls{frb} survey
metric and expected rates as a function of beam sensitivity.  The first side
lobes peak at around $-10$ dB and provide a significant increase in sky coverage
compared to just the primary lobes.

% alfaburst-initial-survey/notebooks/ALFABURST_Derived_FRB_Rates.ipynb

\begin{figure}
    \includegraphics[width=1.0\linewidth]{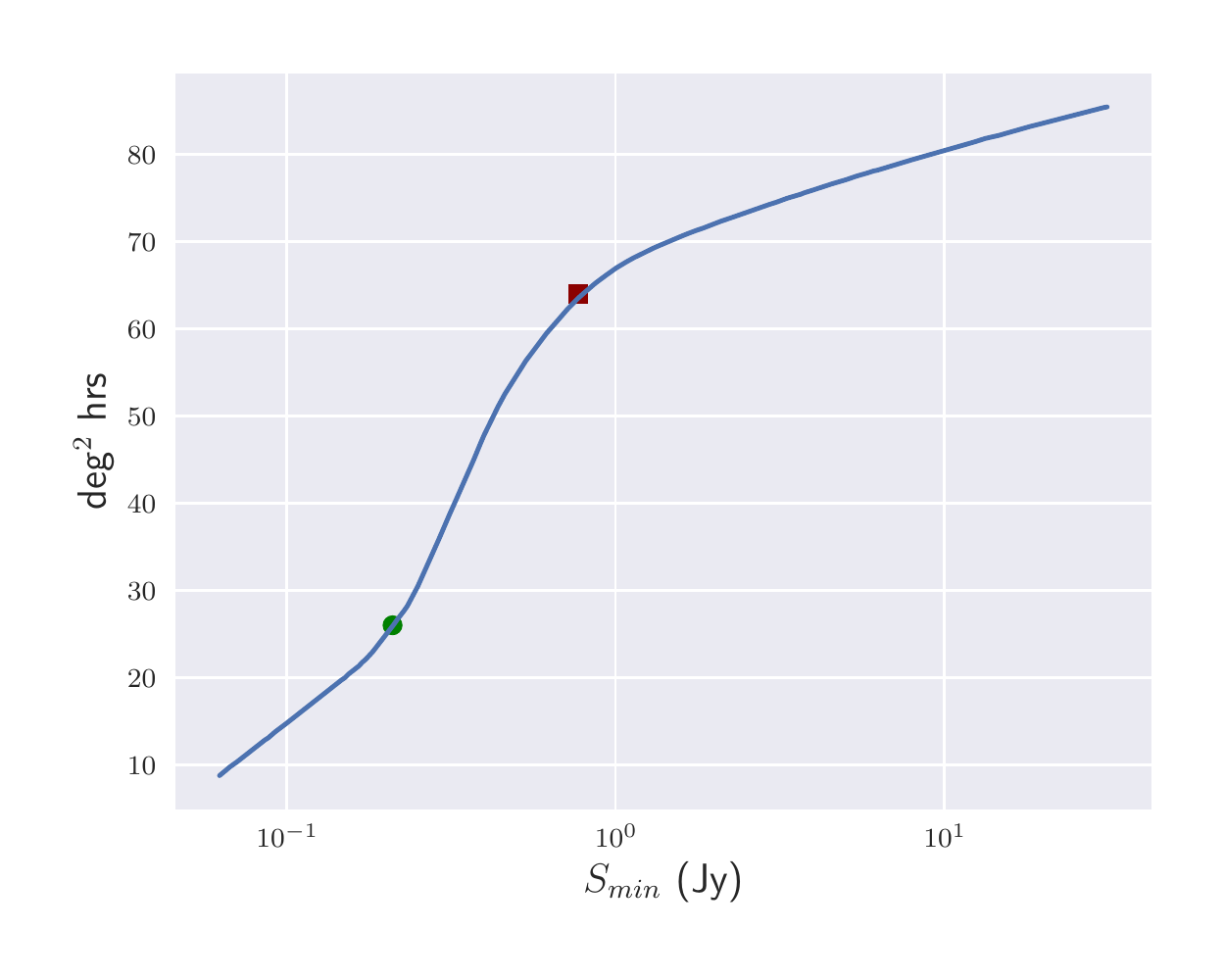}
    \caption{Survey metric as a function of the ALFA receiver minimum
    sensitivity using the ALFA primary and first side lobes. The $-9$ dB point
    (green circle) which is the beginning of the first side lobe sensitivity and
    $-12$~dB point (red square) which is the FWHM of the first side lobe are
    marked.
    }
    \label{fig:survey_metric_sense}
\end{figure}

The total survey metric can be computed as a function of the beam sensitivity by
integrating over the beam (Figure \ref{fig:survey_metric_sense}). We convert the
beam model to units of Jy by assuming that the $-3$ dB point corresponds to the
\gls{fwhm} SEFD of 3~Jy. The survey metric increases to approximately $26 \;
\textrm{deg}^2$ hours by including more of the primary beam beyond the
\gls{fwhm} point. The steep further increase in the survey metric seen in Figure
\ref{fig:survey_metric_sense} arises from including the first side lobes. The
long tail comes from the residual sensitivity by integrating over the remaining
beam. The beam model and polynomial fits to the survey metric curves are
included in the event rate notebooks.
%\textbf{The curve in Figure \ref{fig:survey_metric_sense} can be fit with an
%$8^{\textrm{th}}$-order polynomial $$M_{\textrm{survey}} = \sum_{i=0}^{8} p_i
%(\log S_{\textrm{min}})^i$$
%with coefficents $\{67.12, 26.91, -48.37, 45.87, 30.83, -63.17, 6.64, 23.27, -8.49 \}$}

% alfaburst-initial-survey/notebooks/ALFABURST_Derived_FRB_Rates.ipynb

\begin{figure}
    \includegraphics[width=1.0\linewidth]{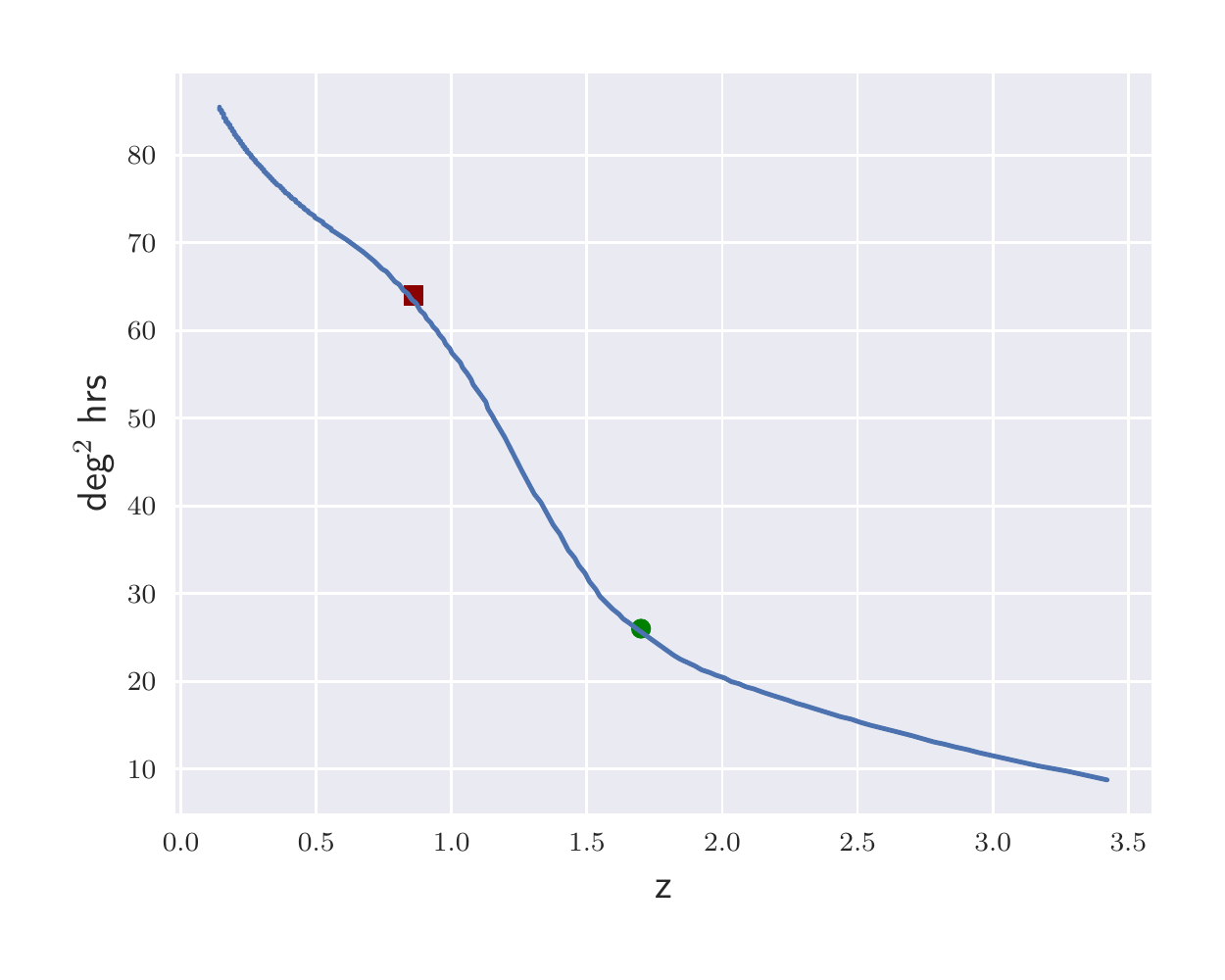}
    \caption{Survey metric as a function of redshift using the standard candle
    model with a flat spectral index ($\alpha=0$) and pulse width of 4 ms. The
    bump out to $z=1.5$ is due to the including the ALFA first side lobes.
    Markers indicate the $-9$ dB (green circle) and $-12$ dB (red square) of the
    ALFA beam.
    }
    \label{fig:full_sefd_z}
\end{figure}

The survey volume is significantly increased by including a large portion of the
beam. It is not possible to put together a figure similar to Figure
\ref{fig:fwhm_sefd_z} when considering the full beam. It is however possible,
under the assumption of flat intrinsic FRB spectra, to compute the maximum
redshift as a function of beam size and sensitivity. Plotting the survey metric
as a function of maximum redshift (Figure \ref{fig:full_sefd_z}) shows how the
full beam model increases the survey metric as a function of redshift. The total
survey volume is computed by integrating over redshift.  Including additional
ALFA side lobes beyond the first side lobes, results in minimal increase in the
survey volume.

% alfaburst-initial-survey/notebooks/ALFABURST_Derived_FRB_Rates.ipynb

The integrated survey volume out to the first side lobe is $5.2 \times
10^6$~Mpc$^3$. The expected number of \glspl{frb} in the survey is 0--2 when
using the galaxy number density and range of $R_{\textrm{FRB}}$ stated above.
Though this event rate is more complex to model, it attempts to provide a more
realistic assessment of the expected detection rates based on the apparent flux
of previously reported \glspl{frb} and a flat spectral index.
%We note that the volume we are sampling is
%biased towards small distances, due to the large sky coverage of the low
%sensitivity outer parts of the beam.

\begin{figure}
    \includegraphics[width=1.0\linewidth]{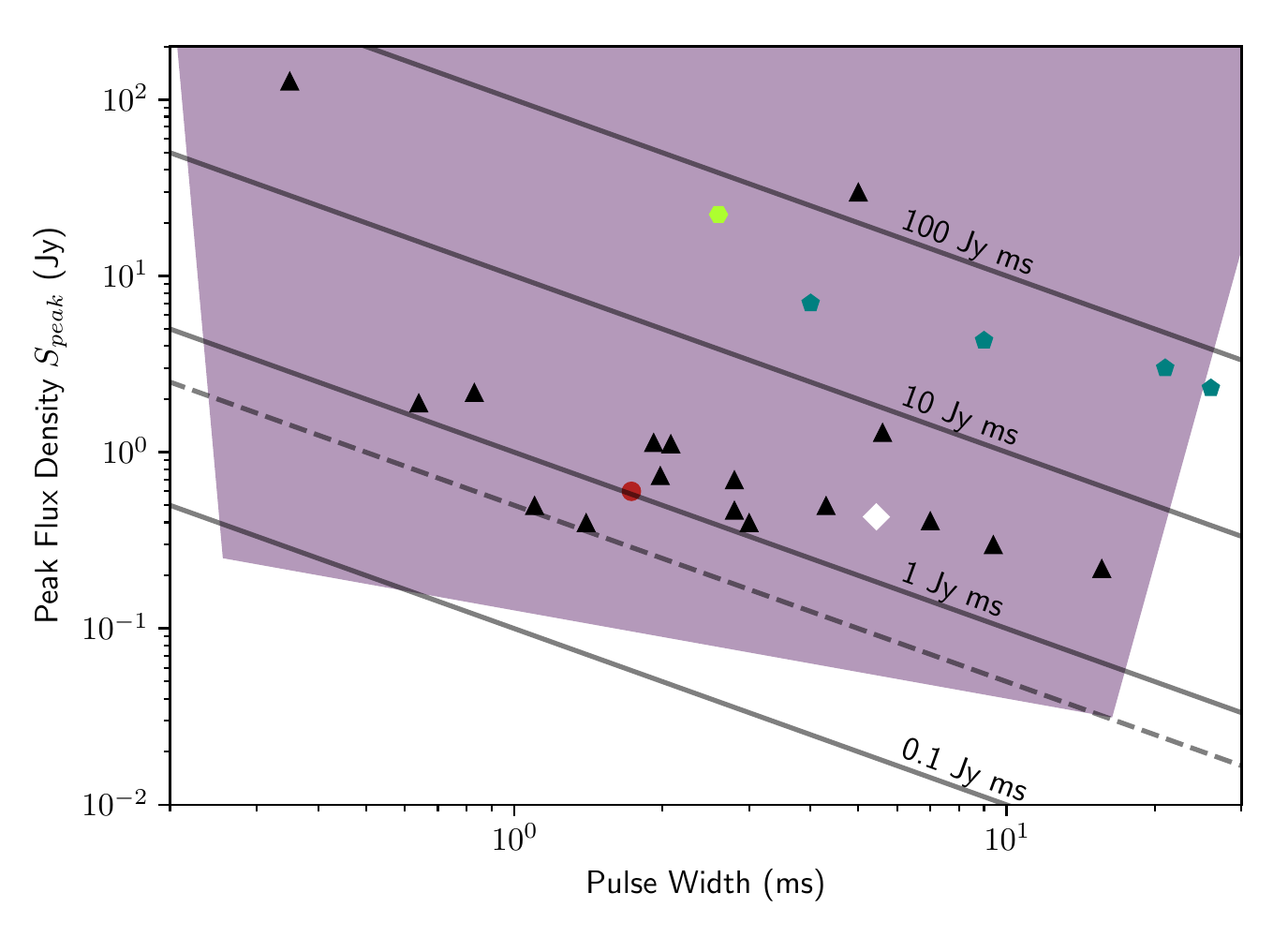}
    \caption{ALFABURST single pulse sensitivity (purple region).
    Previously detected FRBs from Parkes (black triangle), GBT
    (red circle), Arecibo (white diamond), UTMOST (teal pentagon), and ASKAP
    (yellow-green hexagon) are plotted for reference. Line of constant fluence
    (solid) are plotted for reference. The fluence completeness (dashed) is
    0.5~Jy~ms out to pulse widths of 16~ms.
    }
    \label{fig:sensitivity_range}
\end{figure}

Figure \ref{fig:sensitivity_range} shows the ALFABURST sensitivity  based on
pulse width and peak flux, assuming detection at boresight. The ALFABURST
sensitivity region (purple) indicates the survey would be able to detect all
previously reported \glspl{frb}. Bright \glspl{frb} such as FRB150807 and
FRB170827 would be partially clipped by the inline RFI exciser (Section
\ref{sec:rfi_excise}), but they would sill be detected at a high peak S/N.
Additionally, in a multiple beam system a bright \gls{frb} would be
picked up at a lower flux in the side lobes of nearby beams. Recent detections
with UTMOST \citep{2017MNRAS.468.3746C,atel10697,atel10867} indicate that the
parameter space in pulse width should be extended.  FRB170827 has a measured
pulse width of 26~ms. Currently the pipeline decimates in time out to 16~ms. The
pipeline is still sensitive to wider pulses, but at a loss in S/N as indicated
in the slope on the right side of the shaded region of Figure
\ref{fig:sensitivity_range}.  Similarly, the left side of the region is sloped
as ALFABURST is sensitive to bright pulses with widths narrower than $256~\mu$s.

% alfaburst-initial-survey/notebooks/Fluence_Rate.ipynb

%%% EVENT RATES ENDS   %%%

%%% FLUENCE RATE BEGINS %%%

The fluence completeness of the survey \citep{2015MNRAS.447.2852K} is determined
by the minimum detectable fluence at the maximum sampled pulse width in the
survey. ALFABURST has a fluence completeness of $0.5$ Jy ms up to a pulse width
of 16 ms (Figure \ref{fig:sensitivity_range}). All previously reported FRBs are
within this completeness sample except FRB160317 and FRB170827.

%%% FLUENCE RATE ENDS   %%%

\section{Discussion}
\label{sec:discuss}

In addition to the small searched volume, there may be other factors
contributing to our non-detection of \glspl{frb} with the ALFABURST survey. We
derived an expected event rate based on the telescope sensitivity, observing
time, and a standard candle model \citep{2013MNRAS.436L...5L} where the rate of
FRBs per host galaxy is independent of redshift. This is a simple model based on
updates to the empirical event rates from detections in the \gls{htru} survey
\citep{2013Sci...341...53T} by \cite{2016MNRAS.460.3370C}, and assumes that FRBs
are singular events. All of these assumptions are subject to uncertainty. As
shown by recent statistical studies of the Parkes FRBs, there is growing
evidence that they are not  standard candles, and their event rate is redshift
dependent \citep{2016MNRAS.458..708C,ranethesis}.  In addition, the recent
detections of bright, high-DM \glspl{frb} with ASKAP \citep{2017ApJ...841L..12B}
and UTMOST \citep{2017MNRAS.468.3746C,atel10697} also call into question the
assumption that \glspl{frb} are  standard candles.  The repeating nature of
FRB121102 \citep{2016Natur.531..202S} indicates that there could be multiple
classes of \gls{frb} progenitors, or this standard candle model does not
accurately model event rates. The fact that our simple estimate of 0--2
detections so far is broadly consistent with our actual null detection indicates
that our results are not highly sensitive to these assumptions, but that further
ALFABURST observations will begin to probe the variety of currently highly
uncertain features of the FRB population. We discuss these issues, and other
potentially relevant factors, further below.

The limited processing bandwidth of ALFABURST may be a cause of the survey
non-detection. Multiple detected \glspl{frb} show apparent scintillation and
steep spectral indices. It is not possible to differentiate between an apparent
spectral index induced by the beam or an absolute spectral index from the
source. The localization and repeated detections of FRB121102, however, show
there is significant spectral variation, either intrinsic to the source or due
to the intervening medium. Other \glspl{frb} show frequency-dependent structure
which could be due to beam colorization, intrinsic structure, or due to an
intermediate effect. Plasma lenses in the \gls{frb} progenitor host galaxy could
be modulating the pulse amplitude as a function of frequency and time (if the
source repeats) \citep{2017ApJ...842...35C}. This effect introduces an
additional uncertainty in the \gls{frb} rate modeling as the apparent spectral
indices of detected \glspl{frb} may not be intrinsic. Thus, the observed
frequency structure in an \gls{frb} (repeating or not) would be dependent on
multiple factors including observing frequency, bandwidth, epoch, and even sky
direction. If an \gls{frb} did occur in the field of view of the telescope while
ALFABURST was in operation we could have been unlucky, scintillation or lensing
having caused the pulse in the band to go below the detection threshold.
Assuming no scintillation or lensing, an increase to the full \gls{alfa} band
would result in a $\sqrt{6}$ increase in sensitivity compared to what we
currently have. But, also important is a more complete sampling of the frequency
space if these effects are modulating the pulse.

\cite{2015MNRAS.451.3278M} conclude that the apparent deficit of \glspl{frb} at
low Galactic latitudes is due to diffractive interstellar scintillation. Their
model shows that the true event rate is a factor of $\sim 4$ lower than the rate
reported in \cite{2013Sci...341...53T}, which is also the rate used in the
standard candle model \citep{2013MNRAS.436L...5L}. The ALFABURST survey is
evenly split across high and low Galactic latitudes.  \cite{2015MNRAS.451.3278M}
predict that the increase in sensitivity of Arecibo compared to Parkes should
result in a factor of 14 increase in detections, assuming a similar bandwidth
($\sim 300$ MHz). Accounting for the smaller bandwidth of ALFABURST means there
should still be a factor of a few increase in rates. This non-detection result
indicates that the \cite{2015MNRAS.451.3278M} flux density distribution is not
as steep as predicted, or that the source count distribution begins to
flatten below the Parkes sensitivity threshold \citep[for further discussion on
FRB source counts, see][]{2017arXiv171011493M}.

The sensitivity of Arecibo allows the ALFABURST survey to probe a search volume
out to higher redshifts than other surveys. Our number estimates have assumed
that the density of sources per unit co-moving volume is constant.  If FRBs
are standard candles, and that there is a peak similar to the star formation
rate around $z=2$ \citep{2014ARA&A..52..415M} than the expected event rate that
our deeper ALFABURST survey probes would actually be higher than our simple
estimates.  \citet{2016MNRAS.458..708C} and \citet{ranethesis} show that a
larger sample of FRBs in the Parkes surveys is currently required in order to
distinguish between a constant density versus a redshift dependent model.
Neglecting other factors that might hinder detection, and keeping in mind
the standard candle assumption, our null result suggests that the density of
FRBs per unit co-moving volume does not change substantially.

If \glspl{frb} are inherently not flat-spectrum sources, then their fluxes will
be modified substantially: a steeper negative spectrum population would be
harder to detect, while a rising spectrum population would be more readily
detectable.  \cite{2017arXiv170507553L} report FRB121102 to be band limited
during simultaneous observation campaigns using multiple telescopes to cover a
broad range of the radio band. \cite{atel10675} observed 15 pulses from
FRB121102 across the 4-8 GHz band and reported spectral variation over a brief
period of time. A high redshift, band-limited \gls{frb}, which ALFABURST is
sensitive to, could be shifted below L-band. Such a pulse would not be detected
with ALFABURST.

\section{Conclusions and Future Work}
\label{sec:future_work}

We have described the implementation and initial operations of a commensal
search for transient dispersed pulses using the ALFA receiver on the Arecibo
telescope. In our observations carried out so far, we have detected 17
previously known pulsars and found one new  high DM transient in the Galactic
Plane.  Follow-up observations of the same will hopefully reveal the true nature
of the source. This serendipitous discovery during a slew shows the importance
of developing commensal backends for transient searches on large radio
telescopes.

No new FRBs were found in our observations to date. This appears to be broadly
consistent with the expectations from a simple model in which FRBs are treated
as flat-spectrum standard candles uniformly distributed per unit co-moving
volume. We expect continued observations with ALFABURST to run commensally with
other ALFA projects, leading to an improvement on the event rate limits of
low-fluence \glspl{frb}. Quadrupling the current time on-sky, for example,
would lead to an expectation of several FRBs and allow us to more
quantitatively test the validity of our assumptions about their underlying
population, especially the rate dependence on redshift.

The current \gls{sps} pipeline is undergoing a significant upgrade. The input
bandwidth is limited to 56 MHz of the full 336 MHz digital band due to IO
limitations. A new pipeline developed for \gls{ska} \gls{nip} will be used to
process the full \gls{alfa} band.  This will increase sensitivity, and improve
detection rates for scintillating or lensed \glspl{frb}.  An improved version
of the real-time \gls{rfi} exciser is currently being developed and will be
deployed to reduce the false detection rate. The post-processing classifier and
prioritizer model is being updated to make use of an auto-encoder to select
deep features and auto-generate classes. This will allow for an improved
follow-up and analysis cycle.

Over the time period ALFABURST has been active, the use of \gls{alfa} has
decreased as a number of surveys carried out with it have come to an end. We
are currently generalizing the \gls{alfa} specific \gls{sps} pipeline to be
used when other feeds are active. The results from this study would increase
our survey time, and sample a larger portion of frequency space. 

\section*{ACKNOWLEDGEMENTS}

We thank the referee for constructive comments on the manuscript. ALFABURST
activities are supported by a National Science Foundation (NSF) award
AST-1616042.  MAM was supported by NSF award number AST-1517003. MPS and DRL
were supported by NSF award number OIA-1458952.  A.K., J.C., G.F. would like to
thank the Leverhulme Trust for supporting this work.  G.F., D.M, A.S.
acknowledges support from the Breakthrough Listen Initiative. Breakthrough
Listen is managed by the Breakthrough Initiatives, sponsored by the Breakthrough
Prize Foundation\footnote{breakthroughinitiatives.org}.
 
\bibliographystyle{mnras}
\bibliography{alfaburst.bib} 

\bsp	% typesetting comment
\label{lastpage}
\end{document}